\documentstyle[11pt]{article}
\begin{document}

\title{{\bf Dispersionless Fermionic KdV}}
\author{J. Barcelos-Neto\\
Instituto de F\'{\i}sica,\\
Universidade Federal do Rio de Janeiro,\\
Rio de Janeiro, RJ 21945-970\\
Brazil\\
\\
and\\
\\
Alin Constandache and Ashok Das \\
Department of Physics and Astronomy, \\
University of Rochester,\\
Rochester, New York, NY 14627-0171\\ 
USA.}
\date{}
\maketitle

\begin{abstract}

We analyze the dispersionless limits of the Kupershmidt equation, the
SUSY KdV-B equation and the SUSY KdV equation. We present the Lax
description for each of these models and bring out various properties
associated with them as well as discuss open questions that
need to be addressed in connection with these models.

\end{abstract}

\vfill\eject
\section{Introduction:}

In the last two decades, integrable models [1-3] have drawn a lot of
attention from various points of view. More recently, however, 
research in a particular class
of integrable models, known as the dispersionless limits of integrable
models, has become quite active. These involve equations of hydrodynamic
type [4-8] and include systems like the
Riemann equation [8-10], the polytropic gas dynamics [8,11], the
chaplygin  gas, the
Born-Infeld model [12-13] etc. These are models which can be obtained from a
\lq\lq classical'' limit [6,14] of integrable models where the dispersive
terms are absent. Thus, the dispersionlesss limit of the KdV equation [15],
for example, corresponds to the Riemann equation and so on. The
interesting thing about these models (and the \lq\lq classical''
limit) is that, if we know the Lax operator description of a given
model, then, the dispersionless limit also has a Lax description, in
terms of a Lax function on the phase space, where the commutator is
replaced by a Poisson bracket. The Lax function can, in fact, be
determined from the Lax operator in a systematic manner [7]. (Of
course, there are also dispersionless models whose integrable,
dispersive counterparts are not known.)

While a lot is known about the dispersionless limits of bosonic theories, 
nothing is known about the fermionic theories (with or without
supersymmetry).  The main difficulty lies in the fact that, in dealing
with fermionic models, we have to deal with a phase space with both
bosonic and fermionic coordinates. The classical fermionic variables are
nilpotent and, consequently, a power series representation that is so
crucial in a Lax description seems to fail. Thus, for example, the
supersymmetric KdV equation [16] is described in terms of a Lax operator
which involves powers of the supercovariant derivative $D$. This is a
fermionic operator whose square equals $\partial$ ($={\partial\over
\partial x}$). On the other hand, a phase
space description of the model in the dispersionless limit would
naively seem to require using a fermionic phase space variable which would give
the same action as $D$ through a Poisson bracket relation. Such an
object, however, would be nilpotent and there cannot be a power series
in this variable. Such difficulties have hindered the understanding of
such fermionic theories so far.

In this paper, we make the first attempt towards understanding
such models. We analyze the dispersionless limits of fermionic KdV
equations which include the Kupershmidt equation [17], the SUSY KdV-B
equation [16,18]
as well as the SUSY KdV equation [16]. All these equations can be thought
of as fermionic extensions (with or without supersymmetry) of the
Riemann equation  which are
integrable (Of course, we only concern ourselves with $N=1$
supersymmetry.). We obtain a \lq\lq classical'' Lax description for
each of these models. However, the Lax functions and the Lax equations
are obtained by brute force (with a lot of hard work) and we do not
yet understand a systematic 
way in which the phase space Lax functions (for the fermionic models)
can be constructed from the corresponding Lax operators. This 
remains an open question. From the Lax description, we construct
all the local conserved quantities for the models in the standard
manner. However, as 
we emphasize in the paper, fermionic models, in particular, the
supersymmetric ones contain nonlocal conserved charges [19-20] as well,
and it is not at all clear how a phase space Lax description can 
generate such quantities from a \lq\lq Trace'' of the Lax
function. This, too, is an interesting open question. While the Lax
description of 
the bosonic models in the dispersionless limit does give the first
Hamiltonian structure from a generalized Gelfand-Dikii bracket [10], as we
discuss, except for the SUSY KdV-B equation, we are unable to obtain the
first Hamiltonian structure from a generalized Gelfand-Dikii
bracket for the other two models. This is in spite of a Dirac analysis
which we have carried 
out for the constrained form of the Lax functions [21]. In the case of SUSY
KdV, this may be understood as signifying that the first Hamiltonian
structure of SUSY KdV vanishes in this \lq\lq classical''
limit. However, for the dispersionless Kupershmidt equation, there
does exist a first Hamiltonian structure and it is unclear how to
obtain this from the Lax description itself. Our paper is organized as
follows. In section {\bf 2}, we recall, very briefly, the essential
features of the dispersionless KdV equation. In section {\bf 3}, we
present the Lax description for the dispersionless Kupershmidt equation
and discuss all its properties in detail. In section {\bf 4}, we discuss the
analogous results for the dispersionless SUSY KdV-B equation. Here,
there are nonlocal conserved charges as well, and we present the
algebra of the charges which takes a particularly simple form. In
section {\bf 5}, we discuss results for the dispersionless SUSY KdV
equation. Here, too, there are nonlocal charges and we present the
algebra of the charges in a closed form. In section {\bf 6}, we
present a short conclusion emphasizing  the open questions within the
context of such models.

\section{Dispersionless Limit of KdV Equation:}

In this section, we will briefly review the known features of the
dispersionless KdV equation [8,10]. As is well known, the KdV equation [15] 
\begin{equation}
{\partial u\over \partial t} = 6 u\,{\partial u\over \partial x} +
{\partial^{3}u\over \partial x^{3}}
\end{equation}
can be described by the Lax equation
\begin{equation}
{\partial L\over \partial t} = 4 [(L^{3/2})_{+}, L]
\end{equation}
where the Lax operator
\begin{equation}
L = \partial^{2} + u(x)
\end{equation}
with $\partial$ representing ${\partial\over \partial x}$, $u$ the KdV
variable and $()_{+}$ denoting the differential part of a
pseudo-differential operator. The third derivative term on the right
hand side of eq. (1) represents a dispersive term and the
dispersionless limit of this equation (namely, the equation where the
dispersive term is absent) is obtained as follows. Let
\begin{equation}
{\partial\over \partial t}\rightarrow \epsilon\,{\partial\over \partial
t},\quad {\partial\over \partial x}\rightarrow \epsilon\,{\partial\over
\partial x} 
\end{equation}
without rescaling the dynamical variable. Then, in the limit,
$\epsilon\rightarrow 0$, the KdV equation reduces to
\begin{equation}
{\partial u\over \partial t} = 6 u\,{\partial u\over \partial x}
\end{equation}
which is the Riemann equation and has no dispersive term. This is
known as the dispersionless limit of the KdV equation and it has a Lax
description as well. Consider the Lax function on the classical phase
space
\begin{equation}
L = p^{2} + u(x)
\end{equation}
where $p$ is the momentum variable of the phase space. Then, with the
canonical Poisson bracket relations, it is easy to check that
\begin{equation}
{\partial L\over \partial t} = - 4 \{(L^{3/2})_{+}, L\}
\end{equation}
with $()_{+}$ now representing terms with non-negative powers of $p$,
gives the Riemann equation or the dispersionless limit of the KdV
equation. Thus, one can think of the dispersionless limit as sort of a
\lq\lq classical'' limit where the Lax operator goes into a Lax
function ($\partial\rightarrow p$) and an operator Lax equation involving a
commutator goes into a Lax equation involving Poisson
brackets. Furthermore, the conserved quantities and the Hamiltonian
structures (at least, the first structure) can be obtained from this
Lax function [10]. (The difference in sign between the operator equation
(2) and the phase space equation (7) results from the fact that
$[\partial,f] = {df\over dx}$ while $\{p,f\} = - {df\over dx}$. This would be
reflected in all the cases that we discuss.)

\section{Dispersionless Limit of Kupershmidt Equation:}

The Kupershmidt equation [17] is a nontrivial fermionic extension of the KdV
equation which is integrable. The dynamical equations, in this case,
involve a bosonic variable $u$ as well as a fermionic variable $\psi$
and  are given by the coupled equations
\begin{eqnarray}
u_{t} & = & u_{xxx} + 6 uu_{x} - 12 \psi\psi_{xx}\nonumber\\
\psi_{t} & = & 4 \psi_{xxx} + 3 u_{x}\psi + 6 u\psi_{x} 
\end{eqnarray}
where the subscripts denote differentiation with respect to the
particular variable. This is not a supersymmertric system, but gives a
nontrivial, coupled boson-fermion system (reducing to the KdV
equation when $\psi=0$) which has a
bi-Hamiltonian structure and is integrable.

The Kupershmidt equation can be described by a Lax equation. Consider
a Lax operator of the form
\begin{equation}
L = \partial^{2} + u + \psi \partial^{-1}\psi
\end{equation}
Unlike the KdV equation, this Lax operator is truly a
pseudo-differential operator involving the fermionic variable
$\psi$. Once again, it can be easily checked that the Lax operator
equation
\begin{equation}
{\partial L\over \partial t} = 4 [(L^{3/2})_{+}, L]
\end{equation}
gives the Kupershmidt equations.

In trying to derive the dispersionless limit of the Kupershmidt
equation, one runs into various problems. First, we note that, under
the scaling $\partial\rightarrow\epsilon\partial$,
\begin{equation}
\partial^{-1} \rightarrow  \epsilon^{-1}\,\partial
\end{equation}
so that in the limit $\epsilon\rightarrow 0$, the term in the Lax
operator containing the fermionic variables would appear to diverge. Second,
if we naively let $\partial\rightarrow p$, then, of course, the
fermionic term in the Lax operator would vanish. An alternate approach
is to recognize that $\partial^{-1}$ is really an operator which can
be taken to the right with the help of the Leibnitz rule, giving an
infinite series of terms in which one can let
$\partial\rightarrow p$ and the  terms involving the fermionic
variables would no longer vanish. However, a short calculation shows that
such a procedure leads to an inconsistent Lax equation. Therefore,
finding a Lax function and a Lax description for 
the dispersionless limit of the Kupershmidt equation genuinely poses a
challenge. 

The solution to this problem comes as follows. Consider the Lax
function
\begin{equation}
L = p^{2} + u - p^{-2}\psi\,\psi_{x}
\end{equation}
Namely, the term containing the fermionic variables corresponds only
to the first nontrivial term in taking the operator $\partial^{-1}$ to the
right and setting $\partial=p$. It is now easy to check that the Lax
equation (once again, note the difference in sign)
\begin{equation}
{\partial L\over \partial t} = - 4 \{(L^{3/2})_{+}, L\}
\end{equation}
gives rise to the equations
\begin{eqnarray}
u_{t} & = & 6 uu_{x} - 12 \psi\psi_{xx}\nonumber\\
\psi_{t} & = & 3 u_{x}\psi + 6 u\psi_{x}
\end{eqnarray}
which indeed represent the dispersionless limit of the Kupershmidt
equation in (8). (It can be checked that the \lq\lq classical'' limit 
involves the scaling $\partial\rightarrow \epsilon\partial$ and
$\psi\rightarrow \epsilon^{-1/2}\,\psi$ without which the fermion
terms would not be present in the boson equation. This is indeed very
different from the clasical limit in bosonic theories. We will see
this in all the fermionic models we describe.) The passage from the Lax
operator description of the Kupershmidt equation to the Lax function
description of the corresponding dispersionless equation clearly is not as
straightforward  as the bosonic KdV equation and it is not at all
clear how the Lax function of eq. (12) could have been deduced from
that of eq. (9).

The existence of a Lax description, as we know, gives, in a simple
way,  many of the interesting properties of the integrable
system. Thus, for example, from the Lax function, we can define
\begin{equation}
H_{n} = {\rm Tr}\,L^{2n+1\over 2} = (-1)^{n+1}\,{{2n+1\over 2}\choose
n+1}\,\int dx\,\left[u^{n+1}-2n(n+1) u^{n-1}\psi\psi_{x}\right]
\end{equation}
where \lq\lq Tr'' represents the integral of the coefficient of the
$p^{-1}$ term in the expression. We can, in fact, easily check, using
the equations of motion, that the $H_{n}$'s are
conserved. Furthermore, these can be identified with the limit
$\epsilon\rightarrow 0$ of the conserved charges for the Kupershmidt
equation under the scaling
\begin{equation}
\partial\rightarrow \epsilon\partial,\quad
\psi\rightarrow\epsilon^{-1/2}\psi
\end{equation}
This is again a manifestation of the necessity for scaling the fermion
variables to obtain the dispersionless limit (without the scaling,
there would be no fermion terms).

The Lax description, for the bosonic KdV equation, of course, gives
the Hamiltonian structures (at least the first one) naturally through a
generalization of the Gelfand-Dikii bracket [10]. A similar analysis
fails in this case, in spite of a careful treatment (Dirac analysis)
of the constrained 
nature of the Lax function [21-22]. The derivation of the Hamiltonian
structure for the Kupershmidt equation from the Lax description,
therefore, remains an open question. However, the Hamiltonian
structures for this system are not hard to obtain directly. In fact,
it can be checked easily that 
\begin{eqnarray}
{\cal D}_{1} & = & \left(\begin{array}{cr}
                          \partial & 0\\
                          0 & -{1\over 4}
                          \end{array}\right)\,\delta(x-y)\nonumber\\
{\cal D}_{2} & = & \left(\begin{array}{cc}
                         \partial u+u\partial & {1\over
                          2}\partial\psi + \psi\partial\\
                       {1\over 2}\psi\partial + \partial\psi & - {1\over 2} u
                         \end{array}\right)\,\delta(x-y)
\end{eqnarray}
give the first two Hamiltonian structures of the system which can also
be derived from the first two Hamiltonian structures of the
Kupershmidt equation under appropriate scaling (see (16)). Once we know the
Hamiltonian structures, it is straightforward to show that the
conserved quantities are in involution. For example, with the first
Hamiltonian structure in (17), we have
\begin{eqnarray}
\{H_{n},H_{m}\}_{1} & = & -2m(n+1)(m+1)(m-1)\int
dx\,(u^{n+m-2}\psi\psi_{x})_{x}\nonumber\\
 & = & 0
\end{eqnarray}
with the usual assumptions on asymptotic fall off of the dynamical
variables. This proves that the model remains integrable in the
dispersionless limit. 

\section{Dispersionless Limit of SUSY KdV-B Equation:}

The supersymmetric KdV-B equations correspond to a trivial
supersymmetrization of the KdV equation [16,18] and are given by
\begin{eqnarray}
u_{t} & = & u_{xxx} + 6 uu_{x}\nonumber\\
\psi_{t} & = & \psi_{xxx} + 6 u\psi_{x}
\end{eqnarray}
This is a set of simple equations where the boson equation does not
depend on the fermionic variable. However, the pair of equations in
eq. (19) are invariant under the supersymmetry transformations
\begin{eqnarray}
\delta\psi & = & \lambda u\nonumber\\
\delta u & = & - \lambda \psi_{x}
\end{eqnarray}
with $\lambda$  a constant Grassmann parameter of the
transformation. It is this particular supersymmetrization of the KdV
equation which shows up in the supersymmetric one matrix model.

The SUSY KdV-B has a Lax description and, being a supersymmetric
system, the proper description for it is in superspace. Let us define a
fermionic superfield
\begin{equation}
\Phi(x,\theta) = \psi(x) + \theta u(x)
\end{equation}
where $\theta$ represents the Grassmann coordinate of the
superspace. Let us  now define a Lax operator
\begin{equation}
L = D^{4} + (D\Phi)
\end{equation}
where the supercovariant derivative is defined as
\begin{equation}
D = {\partial\over \partial\theta} + \theta\,{\partial\over \partial
x},\quad D^{2} = \partial
\end{equation}
It is now straightforward to check that the Lax equation
\begin{equation}
{\partial L\over \partial t} = 4 [(L^{3/2})_{+}, L]
\end{equation}
where $()_{+}$ refers to non-negative powers of $D$, leads to
\begin{equation}
\Phi_{t} = \Phi_{xxx} + 6 (D\Phi)\Phi_{x}
\end{equation}
which contains both  bosonic and fermionic components of the SUSY
KdV-B equation. 

This Lax operator is in many ways reminiscent of the
Lax operator for the KdV equation and finding the dispersionless limit
of this system, therefore, does not pose too much difficulty. However,
there are some interesting features that arise in the case of the
dispersionless equation which we will describe. First, let us note
that although the Lax operator for the SUSY KdV-B system is written in
terms of the supercovariant derivative $D$, we could have written it
equally well in terms of $\partial$ because of the relation between
the two. Keeping this in mind, let us look at the Lax function
\begin{equation}
L = p^{2} + (D\Phi)
\end{equation}
It is then, easy to check that the Lax equation
\begin{equation}
{\partial L\over \partial t} = -4 \{(L^{3/2})_{+}, L\}
\end{equation}
gives the dispersionless limit of the SUSY KdV-B equation, namely,
\begin{equation}
\Phi_{t} = 6 (D\Phi)\Phi_{x}
\end{equation}
which, in components, takes the form
\begin{eqnarray}
u_{t} & = & 6 uu_{x}\nonumber\\
\psi_{t} & = & 6 u\psi_{x}
\end{eqnarray}
These equations are  indeed the dispersionless limits of the SUSY
KdV-B equations in 
(19) and we can think of them as the trivial supersymmetrization
of the Riemann equation.

The Lax description, as we have seen, gives us the conserved
quantities and it is easy to check that
\begin{eqnarray}
H_{n+1} & = & {2^{n+2}\,n!\over (2n+3)!!}\,{\rm sTr}\;L^{2n+3\over
 2}\nonumber\\
 & = & {1\over (n+1)(n+2)}\,\int dz\,(D\Phi)^{n+2}\nonumber\\
 & = & {1\over n+1}\,\int dx\,u^{n+1}\psi_{x}
\end{eqnarray}
are conserved for $n=0,1,2,\cdots$. Here \lq\lq sTr'' stands for the
integration of the coefficient of $p^{-1}$ over the entire superspace
with $dz=dx\,d\theta$. We have also chosen a particular normalization
for later purposes. There are several things to note about these
conserved charges. First, since they are expressed as integrals over
the superspace, they are automatically invariant under supersymmetry
transformations. However, we see that these charges are fermionic in
nature and this suggests that the Hamiltonian structure for the system
should be odd (anti-bracket structure). This can, in fact, be readily
verified, namely, let us define the dual
\begin{equation}
Q = q_{0} + q_{1}p^{-1}
\end{equation}
so that
\begin{equation}
{\rm sTr}\;LQ = \int dz\,q_{1}(D\Phi) = - \int dz\,(Dq_{1})\Phi
\end{equation}
Then, a generalization of the Gelfand-Dikii definition of the first
Hamiltonian structure [10]
\begin{equation}
\{{\rm sTr}\,LQ, {\rm sTr}\,LV\} = {\rm sTr}\,L\,\{Q,V\}
\end{equation}
yields
\begin{equation}
\{\Phi(z), \Phi(z')\} = - 2 \delta(z-z')
\end{equation}
which is, indeed, the correct Hamiltonian structure for the system (up
to normalization) and is an odd structure. This is a special feature of the
SUSY KdV-B system [23] (as well as the SUSY TB-B [24] system).

One way to understand the odd Hamiltonian structure is as follows. Let
us consider the Lagrangian
\begin{eqnarray}
L & = & \int dz\,\left[{1\over 2}\,\Phi\Phi_{t} - {1\over
(n+1)(n+2)}\,(D\Phi)^{n+2}\right]\nonumber\\
 & = & \int dx\,\left[{1\over 2}(u\psi_{t}-\psi u_{t}) - {1\over
n+1}\,u^{n+1}\psi_{x}\right]
\end{eqnarray}
This Lagrangian describes the dispersionless limit of the SUSY KdV-B
hierarchy as its 
Euler-Lagrange equations and is clearly fermionic. It is then, easy to
work out from the special structure of this Lagrangian that there are
constraints in such a theory which modify the canonical even Poisson
brackets to Dirac brackets which are odd.

The first Hamiltonian structure can be written as a $2\times 2$ matrix
in terms of the components as (up to a normalization)
\begin{equation}
{\cal D}_{1} = \left(\begin{array}{cr}
                     0 & -1\\
                     1 & 0
                     \end{array}\right)
\end{equation}
It is also easy to check from the explicit forms of the conserved
quantities in eq. (30) that the recursion operator (function) for the system is
given by
\begin{equation}
R = \left(\begin{array}{cc}
          u & 0\\
          0 & u
          \end{array}\right) = u\cdot I
\end{equation}
so that
\begin{equation}
\left(\begin{array}{c}
       {\delta H_{n+2}\over \delta u}\\
       {\delta H_{n+2}\over \delta \psi}
      \end{array}\right) = R\,\left(\begin{array}{c}
                                    {\delta H_{n+1}\over \delta u}\\
                                    {\delta H_{n+1}\over \delta \psi}
                                    \end{array}\right)
\end{equation}
The recursion operator, then, leads to the hierarchy of Hamiltonian
structures for the system given by
\begin{equation}
{\cal D}_{n+1} = R^{n}\,{\cal D}_{1} = \left(\begin{array}{cr}
                                             0 & -u^{n}\\
                                             u^{n} & 0
                                             \end{array}\right)
\end{equation}
and all of them are indeed odd structures.

In addition to the fermionic conserved charges, $H_{n+1}$, the
dispersionless SUSY KdV-B equation also has bosonic conserved charges. It is
easy to check that the charges 
\begin{eqnarray}
Q_{n+1} & = & \int dz\,\Phi (D\Phi)^{n}\nonumber\\
 & = & \int dx\,(u^{n+1} - n\,u^{n-1}\psi\psi_{x})
\end{eqnarray}
are conserved. Furthermore, these charges are supersymmetric and
can be obtained from the conserved bosonic charges of the SUSY KdV-B
system (which are obtained by taking \lq\lq super Trace'' of multiples
of quartic powers of the Lax operator). However, the charges, in the
present case, cannot be obtained from the Lax function of the
system. This is, in fact, a puzzling general feature of such
systems. Namely, we know that supersymmetric integrable systems
possess nonlocal conserved charges and they arise, say in the case of
SUSY KdV, from the trace of quartic powers of the Lax operator such as
${\rm sTr}\,L^{2n+1\over 4}$. In the case of operators, this is
meaningful because the supercovariant derivative, $D$, is an operator square
root of  $\partial$. In a classical phase space, however, no such
relation exists and it is not clear how to obtain the analog of the
nonlocal charges in such a case. Thus, for example, in the
dispersionless SUSY KdV-B system, it is easy to see that the quantities
\begin{equation}
q_{n} = \int dx\,u^{n} = \int dz\,D^{-1}(D\Phi)^{n}
\end{equation}
are conserved. However, these are nonlocal quantities and it is not
clear how to obtain such conserved quantities from the Lax
function. Let us note that $q_{n}$'s are purely bosonic, and are not
invariant under supersymmetry transformations (even though they can be
written as an integral over superspace, they involve $D^{-1}$ which
leads to  violation of supersymmetry).

Finally, let us note that we have a set of bosonic and fermionic
conserved charges, in this theory, which are invariant under
supersymmetry. Therefore, it is meaningful to
investigate the algebra of these charges. With the first Hamiltonian
structure, for example, we can easily calculate  the algebra which has
the form
\begin{eqnarray}
\{H_{n},H_{m}\} & = & 0\nonumber\\
\{Q_{n},H_{m}\} & = & 0\nonumber\\
\{Q_{n},Q_{m}\} & = & (n-m)(n-1)(m-1)\,H_{n+m-1}
\end{eqnarray}
The first equation simply says that the conserved Hamiltonians are in
involution so that the system is integrable. The second is an
expression of the fact that the $Q_{n}$'s are conserved under all the
flows of the hierarchy. The nontrivial and really interesting one is
the last one which is reminiscent of supersymmetry algebras. The
important thing to remember in this algebra is the fact that $Q_{n}$'s
are bosonic, $H_{n}$'s fermionic and the bracket is odd.

\section{Dispersionless Limit of SUSY KdV Equation:}

The supersymmetric KdV-B equation represents a trivial supersymmetrization of
the KdV equation and hence there was not much difficulty in
taking its dispersionless limit. The $N=1$ supersymmetric KdV equation, on
the other hand, is a case where we expect some challenge in taking the
dispersionless limit, just like the Kupershmidt equation. Let us recall
that the equation [16]
\begin{equation}
\Phi_{t} = (D^{6}\Phi) + 3 D^{2}(\Phi(D\Phi))
\end{equation}
which in components has the form
\begin{eqnarray}
u_{t} & = & u_{xxx} + 6 uu_{x} - 3 \psi\psi_{xx}\nonumber\\
\psi_{t} & = & \psi_{xxx} + 3 (u\psi)_{x}
\end{eqnarray}
represents the $N=1$ supersymmetric KdV equation which is known to be
integrable. This equation is invariant under the supersymmetry
transformations of eq. (20) and has a Lax description of the following
form. Let us consider a Lax operator of the form
\begin{equation}
L = D^{4} + D\Phi
\end{equation}
Then, it is easy to check that (in this case, there is a degeneracy and
the Lax operator $L=D^{4}-\Phi D$ works equally well) the Lax
equation
\begin{equation}
{\partial L\over \partial t} = 4 [(L^{3/2})_{+}, L]
\end{equation}
gives us the SUSY KdV equation. It is here that we see the fermion
nature of the problem coming into play. Unlike the case of SUSY KdV-B
equation, here the Lax operator cannot be written completely in terms
of the bosonic $\partial$. One has to understand how to take the
\lq\lq classical'' limit of the operator $D$. Of course, classically,
we would have a phase space which would involve both bosonic and
fermionic coordinates. If $p$ and $p_{\theta}$ represent the bosonic
and the fermionic momenta respectively, then, we can define a
fermionic function
\begin{equation}
\Pi = - (p_{\theta}+\theta p)
\end{equation}
which would generate  covariant differentiation through [25]
\begin{equation}
\{\Pi, A\} = (DA)
\end{equation}
for any arbitrary superfield $A$, and would also satisfy
\begin{equation}
\{\Pi,\Pi\} = -2 p
\end{equation}
reminiscent of the operator relation (23). However, $\Pi$ is a
fermionic function and, therefore, nilpotent, which is not quite the
behavior of the operator $D$.

Furthermore, if we look at the supersymmetry transformations of
eq. (20), it is clear that  the scaling $\partial\rightarrow
\epsilon \partial$ would lead to (in the limit $\epsilon\rightarrow 0$)
\begin{equation}
\delta\psi = \lambda u,\quad\quad \delta u = 0
\end{equation}
which gives rise to a nilpotent algebra. This is also clear from an analysis of
the supersymmetric algebra, namely, if it is only the generator of spatial
translation which scales, then, the supersymmetry algebra would become
nilpotent. To have supersymmetry as we know it, in the  \lq\lq
classical'' limit, we need, therefore, to scale the Grassmann
coordinates and the fermionic variables as
\begin{equation}
\theta\rightarrow \epsilon^{-1/2}\theta,\quad\quad\psi\rightarrow
\epsilon^{-1/2}\psi
\end{equation}
This is another indication that, unlike the bosonic variables, fermion
variables need to scale for a consistent \lq\lq classical'' limit.

From the structure of the Lax operator in eq. (45), we note that if we
introduce, in  analogy with the KdV equation, the Lax function
\begin{equation}
L = p^{2} - \Phi\Pi
\end{equation}
where $\Pi$ is defined in eq. (47), it is easy to verify that the Lax
equation leads to inconsistencies. Thus, {\it a priori}, it is not
clear how to proceed in finding a Lax description and the
dispersionless limit. A little bit of analysis shows that the
structure of the Lax function, in this case, is likely to be an
infinite series of the form
\begin{equation}
L = p^{2} + \sum_{n=0}^{\infty} (A_{n}+\Pi B_{n})p^{-n}
\end{equation}
where $A_{n}$ and $B_{n}$ are superfields to be determined
recursively. An infinite series with undetermined coefficients, of
course, would not be very useful in studying the properties of the system.

Surprisingly, however, we have found that the finite order Lax function
\begin{equation}
L = p^{2} + {1\over 2}\,(D\Phi) + {p^{-2}\over
16}\,((D\Phi)^{2}-2\Phi\Phi_{x}) - {p^{-4}\over
32}\,\Phi(D\Phi)\Phi_{x}
\end{equation}
leads to the dispersionless limit of the SUSY KdV equation
\begin{equation}
\Phi_{t} = 3 D^{2}(\Phi(D\Phi))
\end{equation}
through the Lax equation
\begin{equation}
{\partial L\over \partial t} = - 4 \{(L^{3/2})_{+}, L\}
\end{equation}
This is quite a nontrivial Lax function and it is interesting to note
that it does  not involve the fermionic
variable $\Pi$. It is described completely in powers of $p$, and it is
not clear how one could have deduced this particular form of the Lax
function from the 
Lax operator of SUSY KdV in eq. (45).

On the other hand, since we have a Lax description of the
dispersionless SUSY KdV, we can determine the conserved quantities of
this theory. It is easy to check that the quantities
\begin{equation}
H_{n} = {\rm Tr}\,L^{2n+1\over 2} = {1\over 2} {{2n+1\over 2}\choose
n+1} \int dz\,\Phi(D\Phi)^{n}
\end{equation}
are conserved under the evolution of the system. It is interesting to
note that, although the conserved charges are obtained as traces, they
can actually be written in a completely supersymmetric manner and they are
bosonic unlike the case in SUSY KdV-B. Unfortunately, the
Gelfand-Dikii brackets do not give a meaningful first Hamiltonian
structure for this system. While this may seem reminiscent of the Kupershmidt
equation, in this case, it is easy to check that the first Hamiltonian
structure of SUSY KdV vanishes under the scaling (with $\epsilon =0$), namely,
in the \lq\lq classical'' limit. On the other hand, it is not hard to
obtain a Hamiltonian structure of the theory directly (as far as we know, it is
only the first Hamiltonian structure which is known so far to be
derivable from the Gelfand-Dikii brackets in the dispersionless limit
of bosonic theories) and has the structure
\begin{equation}
{\cal D} = -{1\over 2}\,\left(3\Phi D^{2}+(D\Phi)D+
2(D^{2}\Phi)\right)\delta(z-z') 
\end{equation}
This, indeed, coincides with the \lq\lq classical'' limit of the
second Hamiltonian structure of SUSY KdV.

Once we have the Hamiltonian structure, it is easy to check that
\begin{equation}
\{H_{n},H_{m}\} = \int {\delta H_{n}\over \delta \Phi}\,{\cal
D}\,{\delta H_{m}\over \delta \Phi} = 0
\end{equation}
Namely, all the conserved charges are in involution as we would expect
for an integrable system. In addition to these local conserved
quantities, the dispersionless SUSY KdV also has two sets of nonlocal
conserved quantities, namely, one can check directly that
\begin{eqnarray}
F_{n} & = & \int dz\,(D^{-1}\Phi)^{n}\nonumber\\
G_{n} & = & \int dz\,\Phi(D^{-1}\Phi)^{n}
\end{eqnarray}
are conserved under the evolution. It is worth noting that while
$F_{n}$'s are fermionic, $G_{n}$'s are all bosonic and that
\begin{equation}
G_{0} = 4 H_{0}
\end{equation}
Once again, it is not clear how one can derive such nonlocal charges
from a Lax function description.
The algebra of these charges is tedious, but can be calculated with
the Hamiltonian structure of the system in eq. (58) and has the form
\begin{eqnarray}
\{H_{n},H_{m}\} & = & 0 = \{F_{n},H_{m}\} = \{G_{n},H_{m}\}\nonumber\\
\{F_{n},G_{m}\} & = & 0 = \{G_{n},G_{m}\}\nonumber\\
\{F_{n},F_{m}\} & = & nm\,G_{n+m-2}
\end{eqnarray}
Vanishing of the first three brackets is simply a statement that the
Hamiltonians, $H_{n}$ are in involution and that $F_{n}$ and $G_{n}$
are conserved. However, from the other relations, we see that
$F_{n}$ and $G_{n}$ together define a supersymmetry algebra. In
particular, it is not difficult to check that $F_{1}$ corresponds to the
generator of supersymmetry of the system (up to normalization) with
$G_{0}$ ($=4H_{0}$) representing the Hamiltonian.

\section{Conclusion}

Although a lot is known about the structure of dispersionless limits
of bosonic integrable systems, the dispersionless limits of fermionic
integrable models have proven difficult for a variety of reasons. In
this paper, we have made the first attempt towards understanding such
models. We have studied the dispersionless limits of the
Kupershmidt equation, the SUSY KdV-B equation as well as the SUSY KdV
equation all of which can be thought of as fermionic extensions (with
or without supersymmetry) of the Riemann equation. We have obtained
the  Lax function description for each of these
systems and constructed the Hamiltonians for such systems from
them. However, there is as yet no systematic understanding of how such
Lax functions can be deduced from the Lax operator description
of the original theory. We have presented the Hamiltonian structures,
the nonlocal conserved charges for such systems, as well as the algebra
of conserved charges in  closed form. However, it is not clear how a
Lax function description can give rise to nonlocal conserved
quantities nor is a generalization of the Gelfand-Dikii brackets for
the Kupershmidt and SUSY KdV equations is known. While we have
discussed many interesting  features of such fermionic models, we have
also brought out several interesting open questions that deserve further study.

\section*{Acknowledgments}

One of us (A.D.) would like to thank J.C. Brunelli for collaboration
in the beginning stages of the work and Z. Popowicz for useful
discussions. J.B.-N is supported in part by Conselho Nacional de
Desenvolvimento Cient\'{\i}fico e Tecnol\'{o}gico - CNPq, Brazilian
research agency while A.C. and
A.D. are supported in part by the U.S. Dept. of Energy Grant  DE-FG
02-91ER40685  as well as NSF-INT-9602559.

\end{document}